# PiEEG-16 to Measure 16 EEG Channels with Raspberry Pi for Brain-Computer Interfaces and EEG devices

Ildar Rakhmatulin, PhD, PiEEG

**GitHub**   https://github.com/pieeg-club/PiEEG-16
**Website**   https://pieeg.com/
**YouTube** https://youtu.be/tjCazk2Efqs

**Abstract**.
his article introduces a cost-effective gateway into the fascinating world of neuroscience: the PIEEG-16, a versatile shield for RaspberryPi designed to measure 16 channels of various biosignals, including EEG (electroencephalography), EMG (electromyography), and ECG (electrocardiography) without any data transfer over the network (Wi-Fi, Bluetooth) and processing and feature ectraction directly on the Raspberry in real-time. This innovative tool opens up new possibilities for neuroscience research and brain-computer interface experiments. By combining the power of RaspberryPi with specialized biosignal measurement capabilities, the PIEEG-16 represents a significant step forward in democratizing neuroscience research and exploration.

## 1. Introduction

Electroencephalography (EEG) is one of the most straightforward and accessible methods for capturing brain signals. This technique involves recording electrical impulses from various brain regions using electrodes attached to the scalp. EEG signals are utilized for a wide range of purposes, including both research and clinical applications, and the methods and areas of application are continually expanding.

Different types of sensors can be used to record brain signals, including wet and dry electrodes, as well as contact and non-contact types. Wet electrodes are particularly popular due to their low impedance, which typically ranges from 200 kOhm before gel application to 5 kOhm after gel application . Li et al. conducted a detailed review of the advantages and disadvantages of both wet and dry electrodes, highlighting that each type has its own set of benefits and drawbacks. Another article offers a comprehensive review of dry electrodes manufactured between 2010 and 2021.

In our work, we prefer dry electrodes because they simplify the EEG measurement process by eliminating the need for gel application. However, high contact resistance, resulting from insufficient electrical connection at the electrode-scalp interface, can pose a significant challenge during measurement. Addressing this issue is beyond the scope of this article [1].

## 2. Problem statement and motivation

However, these devices can often be challenging for beginners to use. Our goal is to develop a simple, cost-effective device that makes it easy for newcomers to start measuring biodata.

We decided to make a shield for a single board computer for RaspberryPi because in this case it is possible to visualize data and perform feature extraction directly on the device and without transmitting data via Wi-Fi.

The Raspberry Pi has become the industry-standard choice for single-board computers (SBCs) since its launch in 2012. This popularity has led to the emergence of numerous similar devices, such as Banana Pi and Orange Pi. Noting that despite the Raspberry Pi's limitations in computing power compared to devices like the Jetson Nano, it remains one of the most popular SBCs due to its extensive community support and resources available for developers.

## 3. Review of related works

Today, there is a wide range of devices available for recording EEG signals, each with unique features and applications. For instance, Gunawan et al. [2] utilized an Emotiv device for various classification tasks, while Ashby et al. [3] employed the same device for classifying mental visual tasks. Seneviratna et al. [4] introduced a device capable of transmitting data via Bluetooth from seven EEG channels and one audio channel. Tyler et al. [5] introduced a new brain-computer interface (BCI), but their paper lacks detailed technical information, such as component specifications and PCB layout. Some researchers have experimented with using Raspberry Pi for EEG signal acquisition. For example, Dhillon [6] used an MCP3008 ADC to measure EEG signals for diagnosing brain injuries. However, the MCP3008, with its 10-bit resolution, is inadequate for capturing the subtle voltage changes typical of EEG signals, which are usually in the microvolt range. Low-resolution ADCs are generally suitable only with active electrodes, as noted by Apriadi . Our first device PiEEG [7] was created for 8 channels and PiEEG-16 created to measure 16 channels.

## 4. Technical realization

The ADS1299 device by Texas Instruments can be considered the leader among ADCs for EEG measurement tasks. It has been on the market for over ten years and has always been considered one of the best devices on the market. The main difference between this ADC and its competitors is the presence of an internal multiplexer. The capabilities of this ADC and the characteristics of the ADS1299 multiplexer were discussed in detail in our work [8], but this discussion is left outside the scope of this article.

As a helmet/cap, we used the one from the PiEEG website https://pieeg.com/pieeg-shop/ 16 electrodes were installed according to the international electrode placement system "10-20". The general view of the board is shown in Fig. 1.

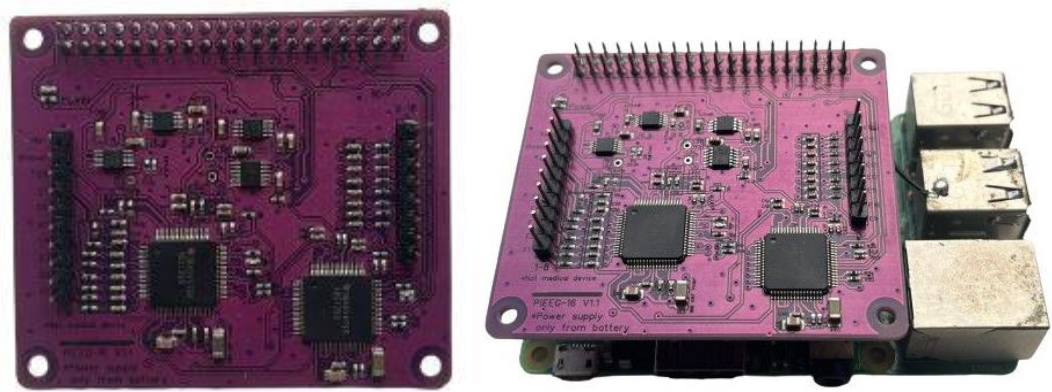

Fig.1. General view of the PiEEG -16 device

During the test, the device was not connected to the electrical source, this was done both for safety and to avoid network interference. A general view of the complete assembly of the device is shown in Fig.2.

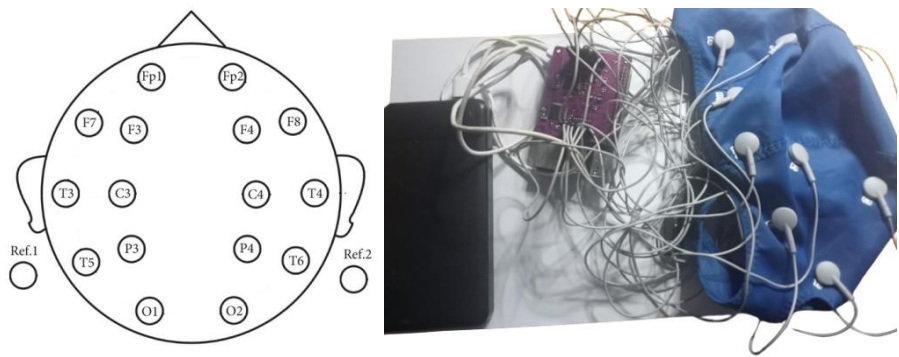

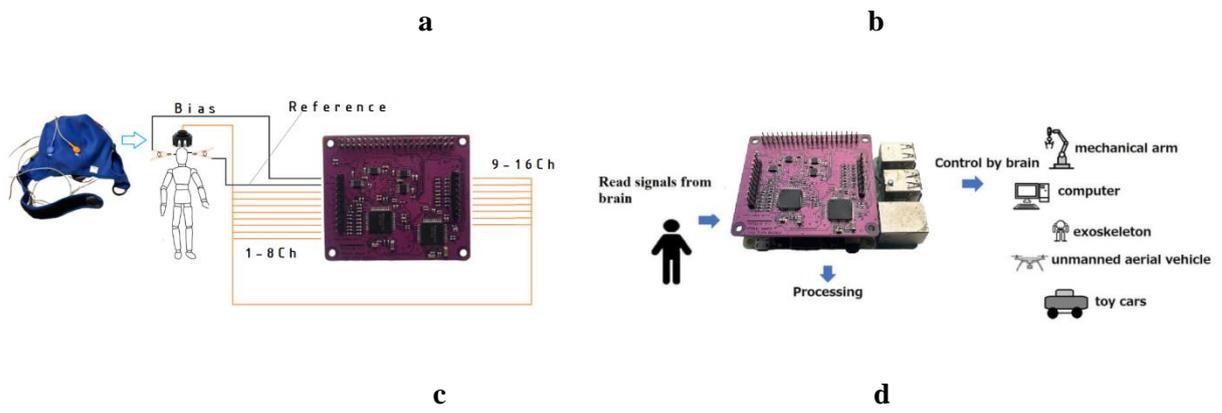

Fig. 2. Genereal View of PiEEG device: a-locations of electrode , b – connections electrodes and device, c – general schematic, d – example of application

### 4.1. Chewing and blinking Artifacts
Test for chewing and blinking artifacts in Fig.3.

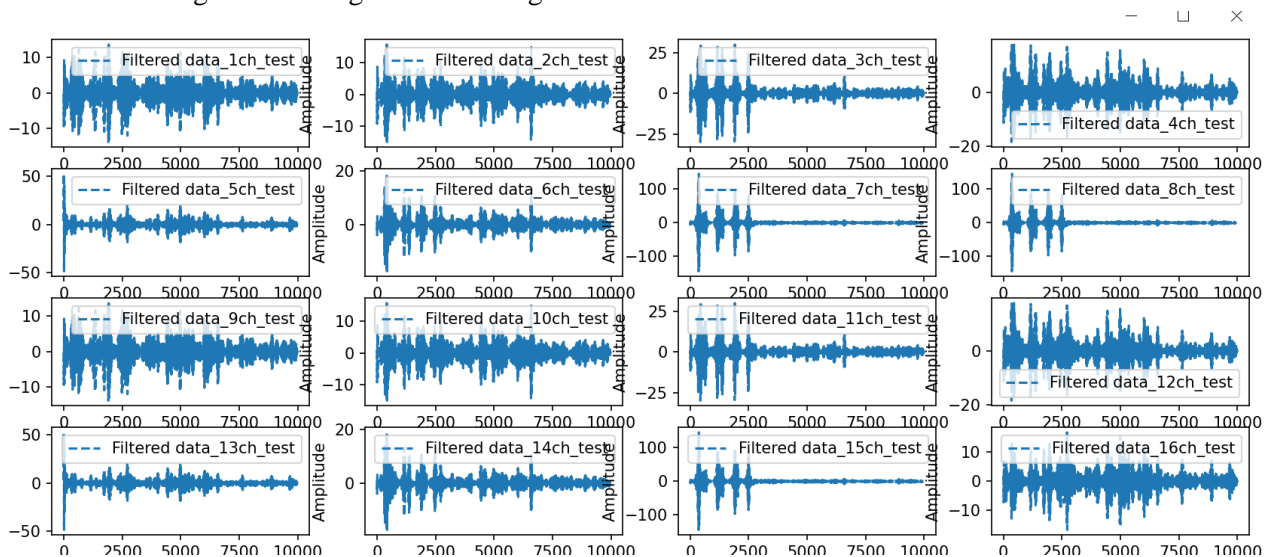

Fig.3. Artifact test. The process of measuring chewing and blinking artifacts using dry electrodes (Fz). Chewing occurred in the following sequence: 4 times, 3 times, 2, and 1 time, and the same for the blinking process. The y-axis is the processed EEG signal after passing filter bands of 1-40 Hz in microvolts and with 250 samples per second (dataset - https://github.com/pieeg-club/PiEEG-16/blob/main/Dataset/2.Chewing_Blinking.xlsx )

### 4.2. Alpha rhythm
Alpha wave (α-wave) brain signal from 7 to 15 Hz with an approximate amplitude selection of 35-65 μV can be used to test the EEG capture system. Generally, alpha waves can be detected in awake people with closed eyes during the relaxation process. We used 16 electrodes for the test. Each time we measured the EEG signal - 5 seconds for closed eyes and 5 seconds for open eyes. As expected, we found an increase in the amplitude of the EEG signals in the frequency band from 9 to 13 Hz. Also, as expected, we found a weakening of the alpha activity for the time when the eyes were open. These results confirm that the device was designed and works correctly. Fig. 4.

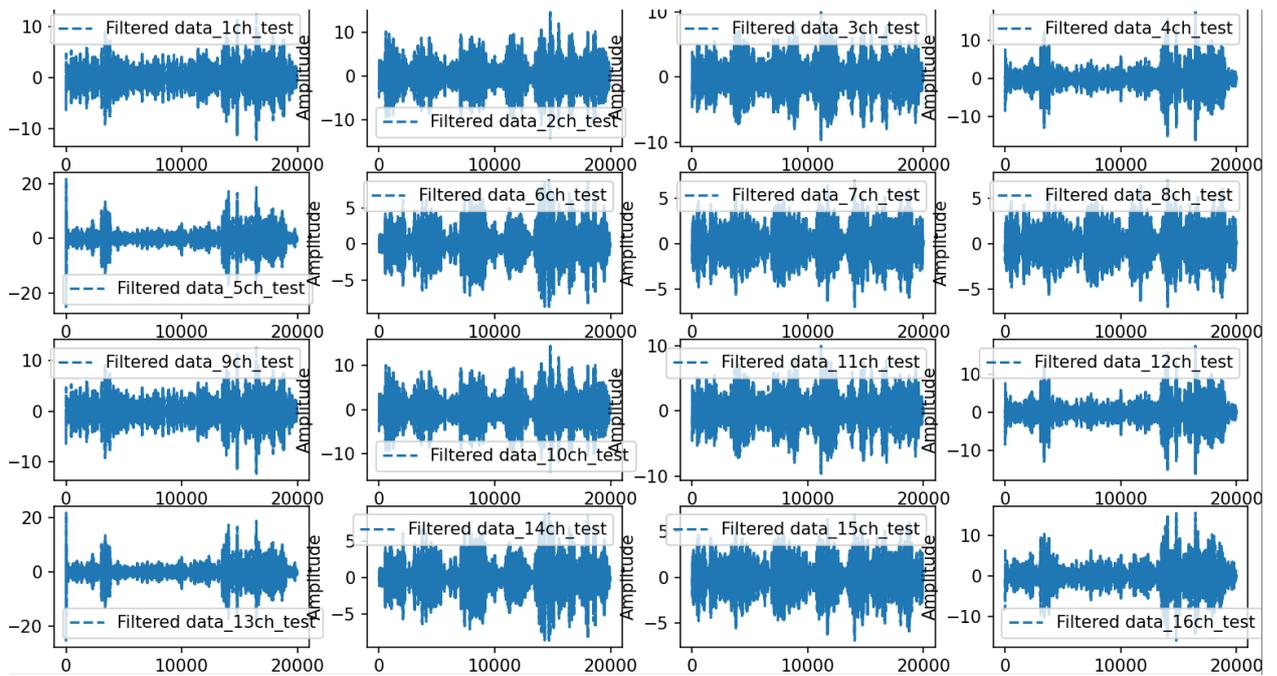

Fig.4. Alpha test. The process of recording an EEG signal from an electrode (Fz) with eyes open and closed. The y-axis is the processed EEG signal after passing filter bands of 8-12Hz in microvolts and with 250 samples per second (dataset - https://github.com/pieeg-club/PiEEG-16/blob/main/Dataset/3.Alpha_test.xlsx)

## 5. Python SDK

For Python was created SDK to collect and visualization EEG data. Example of data visualization in real-time via SDK GUI in Fig.5

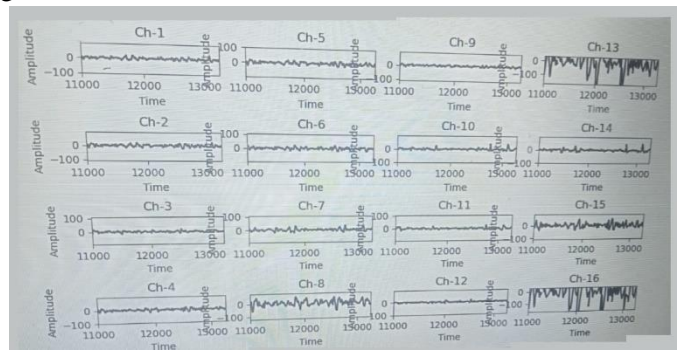

Fig. 5. Real-time, with one second update for 16 channels for PiEEG-16

Sources for SDK https://github.com/pieeg-club/PiEEG-16

## 5. Conclusion and Discussion

In this paper, a new design of device used to measure 16 EEG signals is proposed. One shield unit can simultaneously measure 16 EEG signals. A distinctive feature of the presented device is its low cost while maintaining high accuracy, which was proven in this paper. A promising further direction for this research is its practical use as a brain-computer interface (BCI). Since this work provides high-quality data transfer between the ADC and the processor without time delays in the process, the proposed device can be used in real-time conditions. BCI is used in several areas today, not least helping people with disabilities. But the corresponding devices today require the use of neural networks to establish correlations in the signals and therefore must use a large amount of data to train these networks. However, even for this study, it is better to use compact, inexpensive devices that can operate autonomously for long periods of time. Raspberry PI is very popular, so we hope that this paper will interest many researchers, with which we hope to collect a large data set. To achieve this, we have made this project open source.

CONFLICTS OF INTEREST: NONE

FUNDING: NONE